\documentclass[conference]{IEEEtran}
\IEEEoverridecommandlockouts
\usepackage{cite}
\usepackage{amsmath,amssymb,amsfonts,amsthm}
\usepackage{graphicx}
\usepackage[compatibility=false]{caption}
\usepackage{subcaption}
\usepackage{algorithm}
\usepackage{algpseudocode}
\usepackage{fancyhdr}
\usepackage{graphicx}
\usepackage{mathtools}
\usepackage{booktabs}
\usepackage{color}

\def\BibTeX{{\rm B\kern-.05em{\sc i\kern-.025em b}\kern-.08em
    T\kern-.1667em\lower.7ex\hbox{E}\kern-.125emX}}
\begin{document}

\title{Optimized Sparse Network Coverage via L1-norm Minimization}
\author{\IEEEauthorblockN{Souvik Paul$^{\dag\ddag}$, Iv\'{a}n Alexander Morales Sandoval$^\ddag$ and Giuseppe Thadeu Freitas de Abreu$^\ddag$}
\IEEEauthorblockA{ $^\dag$\textit{Lenovo Deutchland GmbH, Germany. Email: spaul5@lenovo.com} \\
$^\ddag$\textit{Constructor University, Germany. Emails: [soupaul,imorales,gabreu]@constructor.university}}} 

\maketitle

\begin{abstract}
The selection of nodes that can serve as cluster heads, local sinks and gateways is a critical challenge in distributed sensor and communication networks.
This paper presents a novel framework for identifying a minimal set of nexus nodes to ensure full network coverage while minimizing cost.
By formulating the problem as a convex relaxation of the NP-hard set cover problem, we integrate the graph theoretic centrality measures of node degree and betweenness centrality into a cost function optimized via a relaxed $\ell_1$-norm minimization.
The proposed approach is applicable to static and dynamic network scenarios and does not require location or distance estimation.
Through simulations across various graph models and dynamic conditions, it is shown that the method achieves faster execution times (lower complexity) and competitive sparsity compared to classical greedy and genetic algorithms (GA), offering a robust, distributed, and cost-efficient node selection solution.
\end{abstract}


\section{Introduction}

Decentralized sensor networks are foundational to modern intelligent systems spanning diverse domains such as smart agriculture, healthcare, environmental monitoring, industrial automation, and intelligent transportation systems \cite{11008746}.
In these networks, data aggregation, routing control, and local orchestration rely on specialized nodes that coordinate sensing and communication among distributed devices and their corresponding services.
These specialized nodes, hereafter referred to as \textit{\textbf{nexus nodes}} in this work, are capable of taking on roles such as data aggregators, routing intermediaries, and decision-makers for orchestration.

In practice, a network's design, capabilities and goals vary widely: (a) In sensor-centric deployments, individual nodes monitor local phenomena (e.g., temperature, vibration, pollution) and transmit readings to nexus nodes for processing \cite{8653909}.
For instance, in smart agriculture, moisture sensors distributed across a field relay data to central controllers that optimize irrigation schedules \cite{akyildiz2002survey}.
(b) In hierarchical network orchestration, nexus nodes act as \textit{gateways} aggregating data from lower-tier sensor clusters and interfacing with cloud infrastructure.
Such gateways are typical in industrial IoT deployments where time-sensitive data from machine sensors must be relayed efficiently \cite{gubbi2012internetthingsiotvision}.
(c) In peer-to-peer mesh communication networks, nexus nodes operate as \textit{cluster heads}, dynamically organizing neighboring nodes into local groups to enable scalable, resilient routing protocols, etc.
Applications include search-and-rescue missions or surveillance where direct line-of-sight communication is not guaranteed \cite{10.1016/j.comcom.2007.05.024}.

A fundamental challenge across all these architectures is the efficient selection of nexus nodes under energy, bandwidth, and computational constraints.
Poorly chosen nodes can lead to excessive energy drain, redundant communication, or coverage gaps, especially in large-scale, dynamic networks where node failures or mobility are common. 
Traditionally, cluster head or gateway selection is tackled using heuristics based on node degree, residual energy, or proximity metrics \cite{926982,10.5555/2212318.2212343}.
While effective to some extent, these methods either rely on static topology assumptions or incur high overhead to maintain up-to-date node states.

To overcome these limitations, we propose a robust and distributed node selection framework based purely on \textit{connectivity structure}, avoiding the need for distance or positioning measurements.
We formulate the nexus node selection problem as a graph-structured sparse set cover problem (where the objective is to cover the universe using the minimum number of node subsets), integrating graph-theoretic centrality measures (e.g., betweenness, degree) into the optimization cost.
The classical weighted set cover problem is an NP-hard optimization problem which can be solved by a greedy approach that iterates over the sets and selects the one that maximizes the coverage per unit cost, making locally optimal decisions at each step \cite{slavik1996tight,hassin2005better,chvatal1979greedy}.
While more recently, genetic algorithms have also been shown to be effective to optimize selection or placement of such nodes in the graph-network \cite{rai2012multi,larhlimi2025genetic,5586423}.
However, both of the former solutions suffer from slow or premature convergence, especially when the solution space is large or highly multimodal (many local optimum), which may lead to suboptimal covers that either violate constraints or incur excessive costs and complexity \cite{beasley1996genetic,a16070321}

%
In this work, a convex relaxation using an $\ell_1$-norm minimization approach along with a novel cost derived from betweenness centrality and node degree enables the identification of minimal-cost nexus nodes sets that guarantee full network coverage with reduced cost both in a static as well as dynamic environment.
Our key contributions can be summarized as follows
\begin{itemize}
\item A connectivity-based formulation to perform nexus node selection across various network architectures and applications, including sensor, gateway-based, and peer to peer mesh networks.
\item A novel cost function that leverages node centralities to prioritize topologically important nodes while balancing coverage.
\item An efficient convex optimization framework for nexus node selection in static and dynamic settings.
\end{itemize}

\section{System Model}
We consider a network represented by an undirected and connected graph $G = (V, E)$, where $V = \{1, 2, \ldots, n\}$ is the set of nodes and edges $ E \subseteq V \times V $ denotes the set of communication links between the nodes.
A link $(u,v) \in E$ implies that nodes $u$ and $v$ can directly communicate with each other. The network is assumed to be fully connected, meaning there exists a path between any pair of nodes. 
Each node $ v \in V $ can either be selected as a \emph{nexus node} or be covered by a neighboring nexus node.

The objective is to ensure that every node is in direct proximity of at least one nexus node, thus guaranteeing full access to the network to the additional functionality provided by the designated nexus nodes.
To compactly represent this relationship, the coverage  matrix $\mathbf{A}$ is introduced, which is simply defined as
\begin{equation}
    \mathbf{A} = \mathbf{A}_{adj} + \mathbf{I}_n,
\end{equation}
where $\mathbf{A}_{adj}$ is the well-known adjacency matrix, and $\mathbf{I}_n$ is an $n \times n$ identity matrix.

Each entry $ [\mathbf{A} ]_{uv} = 1 $ if node $ v $ can cover node $ u $; that is, either $ u = v $ (self-coverage) or $ (u,v) \in E $ (neighbor coverage). Let $ \mathbf{x} \in \{0,1\}^n $ be a binary vector representing the selection of nexus nodes, and $x_v$ each one of its components.
 \begin{equation}
x_v =
\begin{cases}
1, & \text{if node } v \text{ is selected as a nexus node}, \\
0, & \text{otherwise}.
\end{cases}
\end{equation}

The product  $ \mathbf{A} \mathbf{x} $  gives a vector in $ \mathbb{R}^n $ where each entry quantifies how many nexus nodes cover each node $ u \in V $.
The constraint that all nodes must be covered can now be written succinctly as
\begin{equation}
    \mathbf{A} \mathbf{x} \ge \mathbf{1},
\end{equation}
where $ \mathbf{1} \in \mathbb{R}^n $ is the all-ones vector. This formulation is equivalent to a \emph{set cover problem} defined over graph neighborhoods. 

\section{Proposed Nexus Nodes selection scheme}

\subsection{Problem Formulation}
Given the above, we propose a cost function which is able to capture both the local connectivity and global importance of each node.
Specifically, we consider betweenness centrality, which quantifies the extent to which a node lies on the shortest paths between other nodes and is given by \cite{freeman1977centrality}
\begin{equation} \label{eq:centrality}
    C_B(v) = \sum_{s \neq v \neq t} \frac{\sigma_{st}(v)}{\sigma_{st}},
\end{equation}
where $ \sigma_{st} $ is the total number of shortest paths from node $ s $ to node $ t $, and $ \sigma_{st}(v) $ is the number of those paths that pass through node $ v $.

Then, to encourage the activation of highly central and structurally critical nodes, the cost $c_v$ associated with selecting each node, is defined as
\begin{equation} \label{eq:cost}
    c_v \triangleq \frac{1}{\deg(v)^2 \cdot \log \big(1 + C_B(v)+ \varepsilon\big)},
\end{equation}
where the function $\deg(v)$ returns the number of egdes connected to the node $ i $, and $\varepsilon>0$ is a small constant (e.g.\ $\varepsilon=10^{-6}$) added to prevent division by zero and ensure numerical stability of the cost.
Finally, each node's cost can be collected into a single vector $\mathbf{c} \in \mathbb{R}^n$, yielding the following optimization problem

\begin{equation}
\begin{aligned}
\min_{\mathbf{x} \in \{0,1\}^n} &\quad \mathbf{c}^\top \mathbf{x}, \\
\text{s.t.} \quad & \mathbf{A} \mathbf{x} \ge \mathbf{1}.
\end{aligned}
\label{eq:IP}
\end{equation}

Unfortunately, the Problem in \eqref{eq:IP} is known to be non-convex and NP-hard due to it's combinatorial nature and the discrete variables involved\cite{garey1979computers}.
However, it can be relaxed into a linear program by allowing the optimization variable to take continuous values $ \mathbf{x} \in [0,1]^n$, providing a tractable approximation and a foundation for rounding-based algorithms.

The aforementioned relaxed problem can further be interpreted as a weighted $\ell_1$-norm minimization

\begin{equation} 
    \begin{aligned}
    \min_{0 \leq \mathbf{x} \leq 1} & \quad \| \mathbf{c} \odot \mathbf{x} \|_1 \\
    \text{s.t.} & \quad \mathbf{A} \mathbf{x} \geq \mathbf{1},
    \end{aligned}
    \label{eq:optimization}
\end{equation}

where $ \odot $ denotes element-wise multiplication, problem \ref{eq:optimization} is now convex and can be solved efficiently using interior point methods.

\subsection{Algorithm Description and Parameterization}

\subsubsection{Static}

In the static setting described by Algorithm~\ref{alg:static_l1}, we consider a fixed graph $ G = (V, E) $.
Each node $ v \in V $ is assigned a cost defined as inversely proportional to the square of its degree multiplied by the logarithm of one plus its betweenness centrality, with a small constant $ \varepsilon > 0 $ added to ensure numerical stability.

This cost function prioritizes nodes that are both sparsely connected and structurally central. We then solve a convex optimization problem that minimizes the total cost-weighted selection vector, subject to the constraint that every edge in the graph is covered by at least one selected node.
The resulting fractional solution $ \mathbf{x}^* $ is thresholded (e.g., selecting nodes with $ \mathbf{x}^*_i > 0.5  \rightarrow 1$ and $\mathbf{x}^*_i <= 0.5  \rightarrow 0$) to produce a sparse node set $ \mathbf{S} $ that ensures complete edge coverage. 

\subsubsection{Dynamic}

In the dynamic setting, we initialize with a graph $ G_0 = (V_0, E_0) $ and compute a centrality-aware cost for each node using degree and betweenness centrality.
Solving the initial $ \ell_1 $-norm minimization problem yields a fractional solution $ \mathbf{x}^*_0 $, from which the initial node set $ S_0 = \{\, i \mid \mathbf{x}^*_i > \delta \,\} $ is selected.

As new nodes and edges arrive over time, the incidence matrix and cost vector are updated accordingly. At each step, the $ \ell_1 $ solver is warm-started from the previous solution $ \mathbf{x}^*_{\text{prev}} $, setting the selections of newly added nodes to zero.
The updated fractional vector $ \mathbf{x}^*_t $ is thresholded to yield the current cover $ \mathbf{S}_t $. This strategy allows efficient incremental updates to the node set, ensuring full coverage while minimizing redundant computation.

\begin{algorithm}[h]
\caption{Static $\ell_{1}$ based Node Set Cover with Centrality-Based Cost for Nexus Nodes Selection}
\begin{algorithmic}[1]
\small
\Statex \hspace{-4ex}\textbf{Input:} Graph $G=(V,E)$; threshold $\delta$
\Statex \hspace{-4ex}\textbf{Output:} Selected node set $\mathbf{S}$\vspace{-1ex}
\Statex \hspace{-4.4ex}\hrulefill
\State Build matrix $\mathbf{A}\in\{0,1\}^{|V|\times|V|}$;
\State Construct cost vector $\mathbf{c}$ via equations \eqref{eq:cost} and \eqref{eq:centrality};
\State Obtain $\mathbf{x}^*\in\mathbb{R}_{+}^{|V|}$ via solving the optimization problem \eqref{eq:optimization};
\State Threshold and select nodes: $ \mathbf{S} = \{ i \in V \mid \mathbf{x}^*_i > \delta \} $;
\State \Return $\mathbf{S}$.
\end{algorithmic}
\label{alg:static_l1}
\end{algorithm}

\begin{algorithm}[h]
\caption{Dynamic $\ell_1$ Node Set Cover with Warm Start and Centrality-Based Cost}
\begin{algorithmic}[1]
\small
\Statex \hspace{-4ex} \textbf{Input:} Initial graph $G_0=(V_0,E_0)$; node addition schedule $\{\,\Delta V_t\}_{t=1}^T$; threshold $\delta$;
\Statex \hspace{-4ex} \textbf{Output:} Sequence of node covers $\{S_t\}_{t=0}^T$;\vspace{-1ex}
\Statex \hspace{-4.4ex}\hrulefill \vspace{-0.5ex}
\State Construct cost vector $\mathbf{c}$ via equations \eqref{eq:cost} and \eqref{eq:centrality};
\State Build matrix  $\mathbf{A}$ for $G$;
\State Solve the convex problem via equation (\ref{eq:optimization});
\State Threshold and select nodes: $ \mathbf{S}_0 = \{ i \in V \mid \mathbf{x}^*_i > \delta \} $;
\State \Return $\mathbf{S}_0$, store $\mathbf{x}^*_{\text{prev}} \gets \mathbf{x}^*$
\For{$t = 1$ to $T$}
  \State Add $\Delta V_t$ new nodes to $G$;
  \State Update $C_B(i)$ and $\deg(i)$ for all $i \in V(G)$;
  \State Recompute $\mathbf{c}_i$ and incidence matrix $\mathbf{A}$;
  \State Warm-start the $\ell_1$ solver using $\mathbf{x}^*_{\text{prev}}$;
  \State Solve updated LP and obtain new solution $\mathbf{x}^*$;
  \State Threshold and select nodes: $\mathbf{S_t} \gets \{\, i \mid \mathbf{x}_i^* > \delta \,\}$;
  \State \Return $\mathbf{S_t}$;
  \State Update $\mathbf{x}^*_{\text{prev}} \gets \mathbf{x}^*$;
\EndFor
\end{algorithmic}
\label{alg:dynamic_l1_updated}
\end{algorithm}

\section{Performance Evaluation}
\label{sec:performance}

In this section, first, we demonstrate the core innovation of the proposed nexus nodes section scheme and provide the  proof-of-concept for the static (using different network types) and dynamic (adding nodes and edges to the networks) settings.
Next, to measure and validate the practical benefits of the proposed $\ell_{1}$–based optimization, we conduct an extensive comparison against the classical greedy set-cover heuristic \cite{young2016greedy} and genetic algorithm \cite{larhlimi2025genetic,5586423} under two cost profiles: plain node degree and the centrality-driven cost in Equation \eqref{eq:cost}.
The evaluation is split into a \emph{static} where a whole network is processed in one shot and a \emph{dynamic} study where the algorithm is re-invoked after small, incremental topology changes.
Throughout all experiments, we measure two key metrics: (a) execution runtime and (b) selected nexus node count.

\textbf{Implementation and Experimental Configuration}: To solve the weighted $\ell_1$ minimization problem for nexus node selection, we implement a convex optimization framework using the \texttt{cvxpy} library in Python.
The optimization is solved using the \texttt{SCS} (Splitting Conic Solver) with optional warm-start initialization to accelerate convergence in dynamic settings.
All the experiments were conducted on a system  with an AMD Ryzen 7 PRO 6850U CPU and 16 GB RAM running a 64-bit Python environment.

\vfill

\subsection{Static Scenario}
\label{subsec:static}

\begin{figure}[h]
\centering
\begin{subfigure}{0.24\textwidth}
\includegraphics[width=\linewidth]{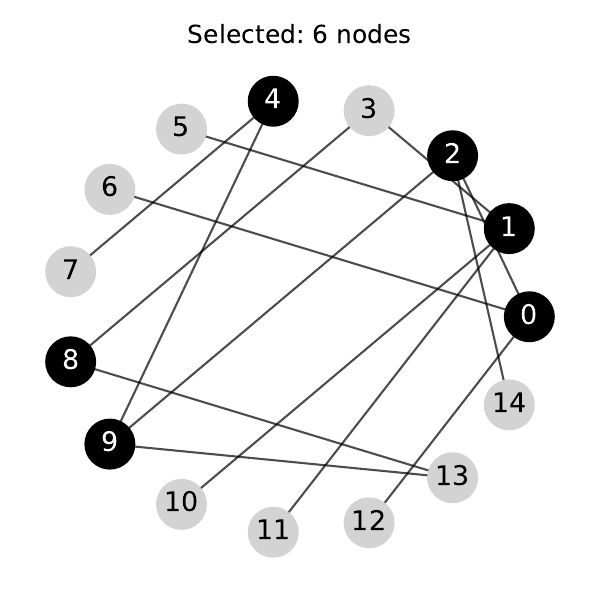}
\caption{Tree Graph}
\label{fig:1}
\end{subfigure}
\begin{subfigure}{0.24\textwidth}
\includegraphics[width=\linewidth]{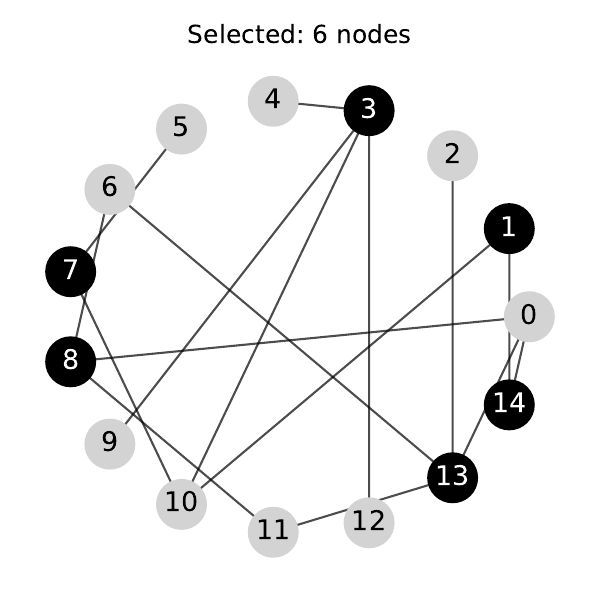}
\caption{Erdős–Rényi Graph}
\label{fig:2}
\end{subfigure}
\begin{subfigure}{0.24\textwidth}
\includegraphics[width=\linewidth]{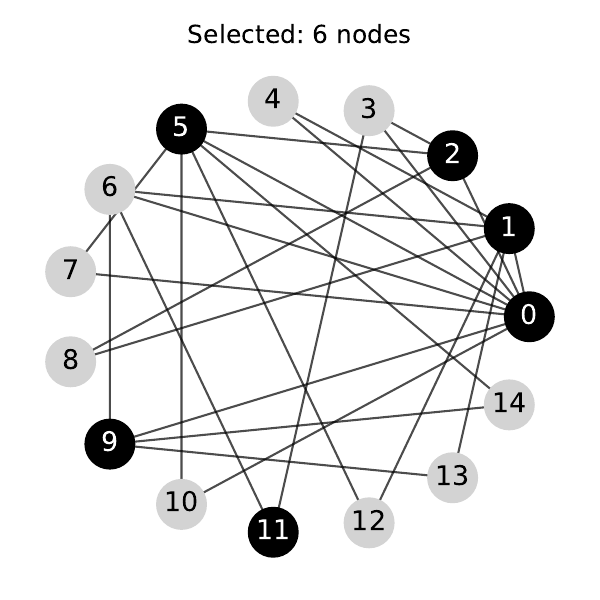}
\caption{Barabási–Albert Graph}
\label{fig:3}
\end{subfigure}
\begin{subfigure}{0.24\textwidth}
\includegraphics[width=\linewidth]{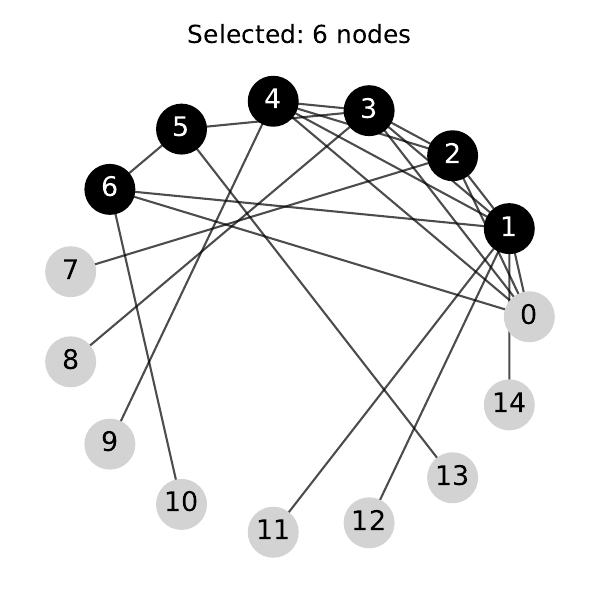}
\caption{Internet Graph}
\label{fig:4}
\end{subfigure}
\caption{Evaluation of the $\ell_{1}$-based nexus node selection with proposed Centrality-based cost on different types of graphs.}
\label{fig:static}
\end{figure}

Figure~\ref{fig:static} displays the effect of the proposed method across various graph types.
Across all graphs generated, namely Tree (uniformly distributed random Prüfer sequences) \cite{diestel2025graph}, Erdős--Rényi (edge probability via binomial distribution) \cite{Erdos2022OnRG}, Barabási--Albert (scale-free via power law degree distributions) \cite{doi:10.1126/science.286.5439.509}, and Internet \cite{5586438}-like topologies with a fixed number of 15 nodes, the algorithm consistently selects 6 nexus nodes for full coverage.
This suggests the robustness of the method across different structural regimes.

\begin{figure}[h]
    \centering
    \includegraphics[width=1\linewidth]{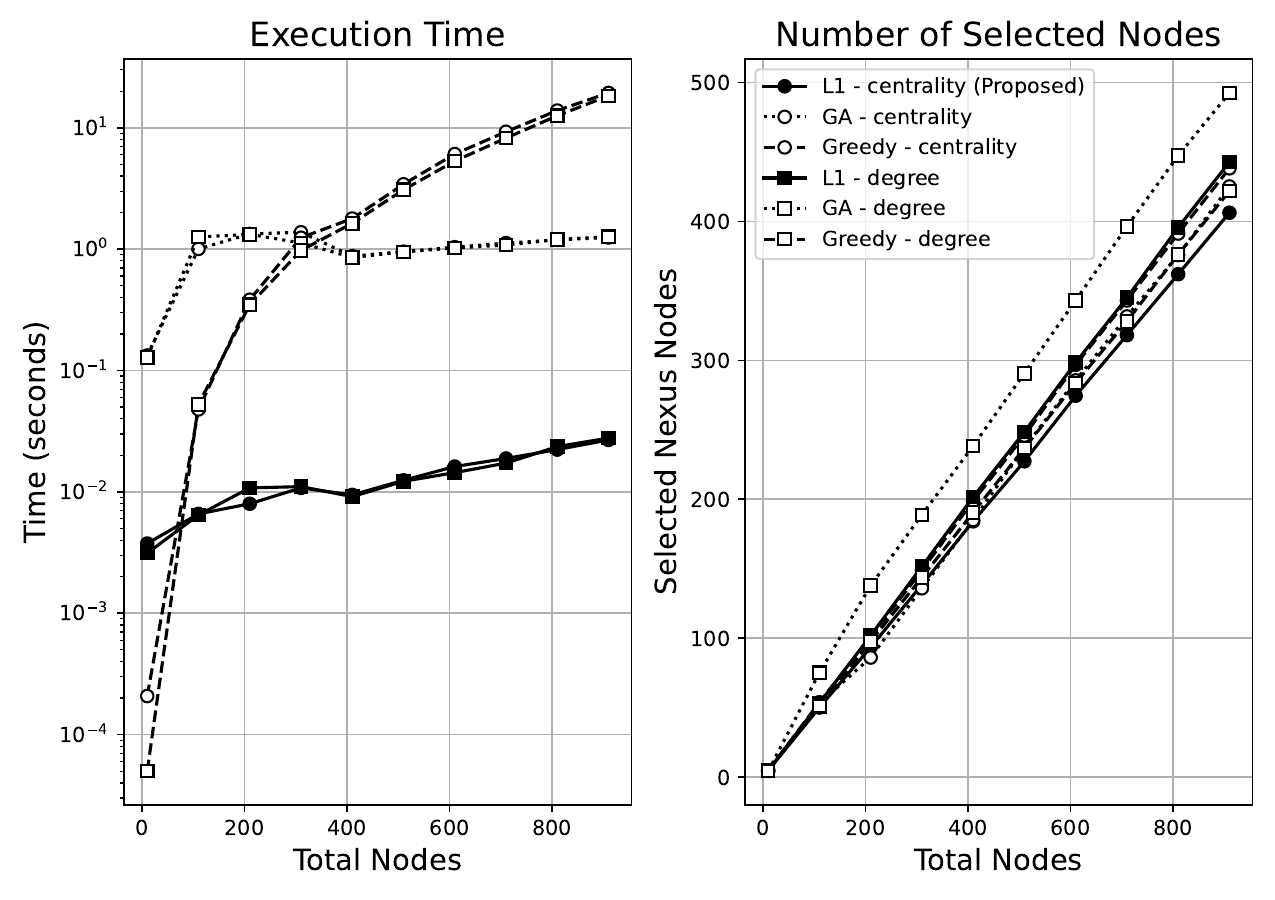}
    \caption{Evaluation of execution-time (left) and number of selected nexus nodes (right) with degree and proposed costs for $\ell_1$-norm optimization, Genetic Algorithm and Greedy.}
    \label{fig:static_output}
    \vspace{-2ex}
\end{figure}

Figure~\ref{fig:static_output} compares the proposed $\ell_1$-based method against Genetic Algorithm and Greedy baselines, using both centrality and degree-based node cost functions.
This is done from 10 to 1000 nodes over an interval of 100 nodes and 20 trials for each of the data points.

The left panel shows execution time (log-scale) as a function of total nodes. The proposed method ($ \ell_1 $-centrality) demonstrates significantly lower execution time compared to GA and Greedy approaches, particularly as the graph size increases.
The right panel reports the number of selected nexus nodes required for full edge coverage. 
The proposed centrality based measure consistently selects fewer nodes than the alternatives, indicating improved efficiency and sparser solutions.
In addition, the mixture of proposed convex relaxation via $\ell_1$-norm and centrality metric is only outperformed for small number of nodes ($< 500$) by the GA-centrality method.
Overall, the proposed approach achieves a favorable balance between runtime and solution compactness.

\vfill

\begin{figure}[t]
\centering
\begin{subfigure}{0.24\textwidth}
\includegraphics[width=\linewidth]{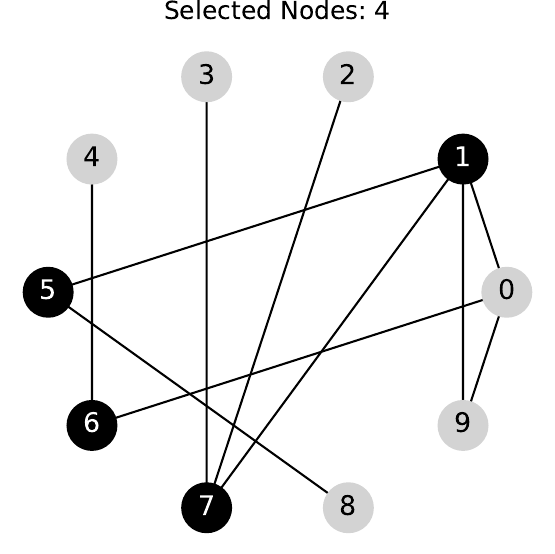}
\caption{Step 1: Selection 4/10 nodes}
\label{fig:snap1}
\end{subfigure}
\begin{subfigure}{0.24\textwidth}
\includegraphics[width=\linewidth]{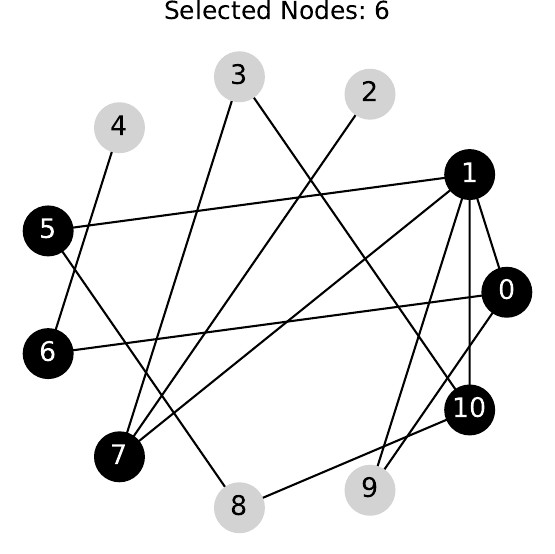}
\caption{Step 2: Selection 6/11 nodes}
\label{fig:snap2}
\end{subfigure}
\begin{subfigure}{0.24\textwidth}
\includegraphics[width=\linewidth]{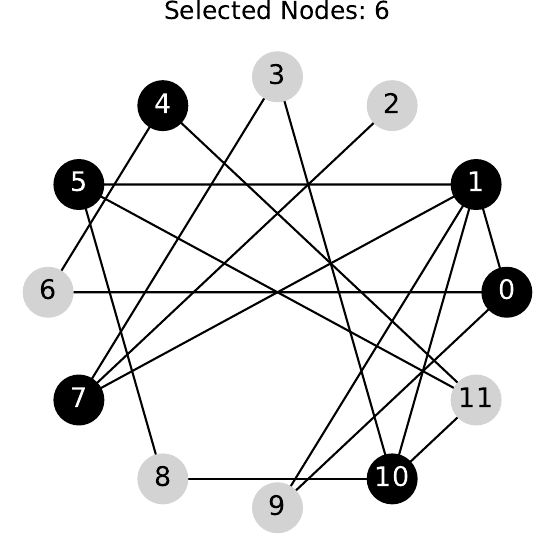}
\caption{Step 3: Selection 6/12 nodes}
\label{fig:snap3}
\end{subfigure}
\begin{subfigure}{0.24\textwidth}
\includegraphics[width=\linewidth]{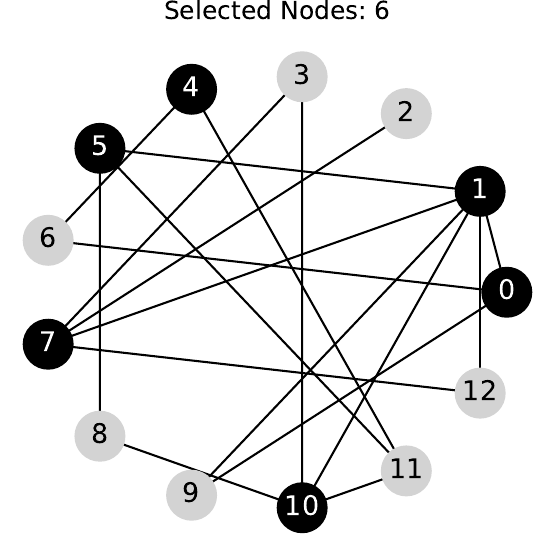}
\caption{Step 4: Selection 6/13 nodes}
\label{fig:snap4}
\end{subfigure}
\caption{Dynamic Selection Progression for $\ell_{1}$-Based Node Set Cover with proposed Centrality-Based Cost.}
\label{fig:all_figures_dynamic}
\end{figure}

\subsection{Dynamic Scenario}\label{subsec:dynamic}


Figures~\ref{fig:snap1}--\ref{fig:snap4} illustrate the dynamic behavior of Algorithm~\ref{alg:dynamic_l1_updated} as the network expands from 10 to 13 nodes.
Each iteration incorporates newly added nodes into the graph and updates centrality measures and the incidence matrix accordingly.
Using the previous solution as a warm start, the convex optimization problem is re-solved  to produce an updated node cover that ensures full coverage.
Initially, the algorithm selects 4 nexus nodes; as the network grows, it adaptively adjusts the selection to maintain a sparse yet effective set of nodes.

\begin{figure}[t]
    \centering
    \includegraphics[width=1\linewidth]{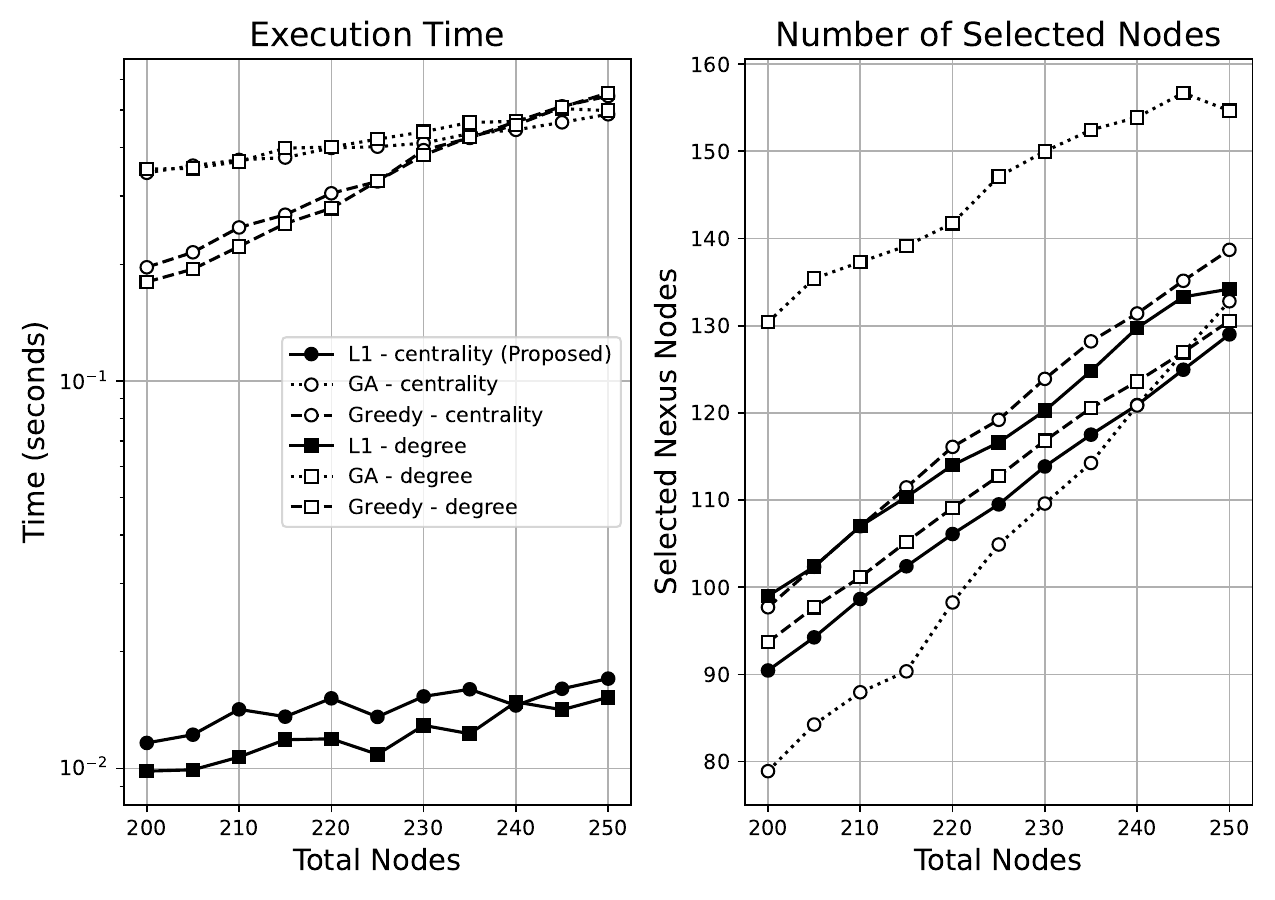}
    \caption{Evaluation of solver performance under incremental node additions with execution time (left) and selected node count (right) for 200–250 node graphs.}
    \label{fig:dynamic_output}
    \vspace{-4ex}
\end{figure}

Figure~\ref{fig:dynamic_output} presents the performance of the proposed $ \ell_1 $-centrality method under a dynamic setting, where the graph evolves over time within a node range of 200–250.
The left panel shows execution time, with the proposed method maintaining a consistently low runtime (around $10^{-2}$ seconds), significantly outperforming GA and Greedy strategies, especially as network size increases. 
The right panel compares the number of selected nexus nodes.
The $ \ell_1 $ centrality method consistently selects fewer nodes than degree-based counterparts and remains competitive with GA-centrality in terms of solution sparsity, while maintaining much lower computational cost. 
This highlights the efficiency and adaptability of our approach in dynamic environments.
\section{Conclusion and Future Work}

This work introduces a convex optimization-based approach to sparse network coverage through centrality-aware nexus node selection.
By leveraging an $\ell_1$-based minimization framework combined with structural properties like node degree and betweenness centrality, our method ensures full coverage with fewer active nodes and reduced computational overhead.
Experimental results demonstrate that the proposed solution consistently outperforms greedy and genetic baselines across static and dynamic topologies, offering faster convergence and more compact solutions.
The framework's adaptability to real-time and evolving networks positions it as a promising tool for future applications, including resilient routing in IoT, agent coordination in distributed AI systems, and quantum network infrastructure.
Future directions include extending the model to support message-passing protocols, social learning mechanisms, and large-scale distributed implementations.

\section*{Acknowledgment}
This work was partially supported in part by the EU’s HE research and innovation program HORIZON-JU-SNS-2022 under the ITRUST6G (Grant No. 101139198) and RIGOUROUS project (Grant No. 101095933) respectively

\bibliographystyle{IEEEtran}
\bibliography{reference}

\end{document}